\documentclass[preprint,12pt]{elsarticle}

\usepackage{graphicx}%
\usepackage{multirow}%
\usepackage{mathrsfs}%
\usepackage[title]{appendix}%
\usepackage{xcolor}%
\usepackage{textcomp}%
\usepackage{manyfoot}%
\usepackage{booktabs}%
\usepackage{algorithm}%
\usepackage{algorithmicx}%
\usepackage{algpseudocode}%
\usepackage{listings}%
\usepackage{hyperref}

\usepackage{dsfont}
\usepackage{siunitx}
\usepackage{multirow}
\usepackage{makecell,multirow}
\usepackage{array}
\usepackage{subcaption}

\usepackage{float}

\usepackage{amssymb}
\usepackage{amsmath}

\begin{document}

\begin{frontmatter}



\title{Hierarchical Representations for Evolving Acyclic Vector Autoregressions (HEAVe)}




\author[adelaide]{Cameron Cornell\corref{cor1}}
\ead{cameron.cornell@adelaide.edu.au}
\cortext[cor1]{Corresponding author}

\author[adelaide]{Lewis Mitchell}
\author[adelaide]{Matthew Roughan}

\affiliation[adelaide]{
    organization={School of Computer and Mathematical Sciences, The University of Adelaide},
    city={Adelaide},
    postcode={5005},
    state={South Australia},
    country={Australia}
}

\begin{abstract}
Causal networks offer an intuitive framework to understand influence structures within time series systems. However, the presence of cycles can obscure dynamic relationships and hinder hierarchical analysis. These networks are typically identified through multivariate predictive modelling, but enforcing acyclic constraints significantly increases computational and analytical complexity. Despite recent advances, there remains a lack of simple, flexible approaches that are easily tailorable to specific problem instances. We propose an evolutionary approach to fitting acyclic vector autoregressive processes and introduces a novel hierarchical representation that directly models structural elements within a time series system.  On simulated datasets, our model retains most of the predictive accuracy of unconstrained models and outperforms permutation-based alternatives. When applied to a dataset of 100 cryptocurrency return series, our method generates acyclic causal networks capturing key structural properties of the unconstrained model. The acyclic networks are approximately sub-graphs of the unconstrained networks, and most of the removed links originate from low-influence nodes. Given the high levels of feature preservation, we conclude that this cryptocurrency price system functions largely hierarchically. Our findings demonstrate a flexible, intuitive approach for identifying hierarchical causal networks in time series systems, with broad applications to fields like econometrics and social network analysis.
\end{abstract}




\end{frontmatter}



\section{Introduction}
Causal modeling has gained significant traction as a subfield of time series analysis, shifting the emphasis from traditional prediction-based paradigms to systems-level inference. Given the natural alignment of time series data with economic analysis, this field is deeply interconnected with econometrics, where policy decisions necessitate a focus on understanding the underlying system dynamics rather than attaining predictive accuracy. For instance, central banks might seek to understand how changes to their interest rates causally influence rates in other countries \cite{RBA_interest}. They may aim to uncover the channels through which these effects propagate—such as trade, investment, or currency markets—and determine the sequence and magnitude of impacts across countries.

While causal analysis often targets a single endpoint variable to evaluate causal effects, here we focus on mutual causation within a system. This naturally gives rise to a causal network, where interdependencies between variables are jointly analyzed to capture system-wide dynamics. The Vector Autoregression (VAR) framework \cite{luet, sims_1980} is the baseline inference method for this kind of multivariate time series analysis. Substantial research efforts have extended its capabilities, with examples like structural and reduced-form VAR, which alter the interpretability of identified effects. 

Restricting the model space to acyclic processes enhances transparency and interpretability. Cycles can obscure how system shocks propagate, making causal analysis challenging. By contrast, a Directed Acyclic Graph (DAG) provides a clear, hierarchical representation of the system, enabling an intuitive understanding of causal flows. Beyond interpretability, the DAG structure unlocks computational advantages by allowing the use of specialized graph algorithms (e.g., topological sorting, shortest path) to analyze and query causal dependencies efficiently. These properties make DAGs particularly valuable for uncovering the structural elements of complex time series systems. Unfortunately, standard VAR models will generally contain a large number of cycles, as they do not impose any constraints on the relationships between variables.

This study introduces \textit{Hierarchically Evolved Acyclic Vector Autoregressions (HEAVe)}, a novel hierarchy-based evolutionary framework for fitting acyclic VAR processes. HEAVe leverages a hierarchical representation to achieve accurate predictions and a robust capacity to identify the structural elements of a time series system.

The main contributions of this study are: 
\begin{itemize}
    \item A novel, hierarchy based evolutionary framework for identifying acyclic Granger causal networks.
    \item A simulation study demonstrating the capacity for HEAVe to produce acyclic VAR models that retain 92\% of the predictive power of unconstrained VAR. The results show clear advantages to the hierarchical approach, with 5.6-8.3\% improved prediction, 6.1-14.1\% improved hierarchy identification and 13.1-19\% improved link classification.
    \item An empirical demonstration of our evolutionary algorithm to a set of cryptocurrency returns, with results indicating that the causal dynamics of this system are well approximated with a hierarchical network. 
\end{itemize}



\noindent Overall, our approach provides an intuitive and flexible framework for imposing acyclic constraints on causal networks, preserving structural elements while enhancing overall interpretability and retaining the majority of predictive power.

\section{Related Work}

The use of graphical structures to capture the joint dependence among random variables is now widely recognized under the framework of probabilistic graphical models. While all fields benefit from understanding joint dynamics, this approach has gained particular prominence in econometrics, where substantial financial incentives drive the study of variable coupling. For instance, the correlation between financial asset returns is a heavily researched topic and can be effectively represented using an undirected graphical model \cite{stat_analysis_of_network}. Extensions to this simple type of financial network include utilizing partial correlations \cite{partial_corr}, or looking at the evolution of correlation structures over time \cite{clustering_and_info, structural_entropy}. 

Sequential dependence structures, such as cross-correlational effects can similarly be represented in a graphical structure. Commonly the contemporaneous relationships are ignored, giving us the Granger-causal network, an approach that has similarly seen econometric applications \cite{billio2011, cornell2023vector, rank2, conf, granger_causal_network}. \footnote{Strictly speaking, such models don't adhere to the standard mathematical definition of a probabilistic graphical model, however, much of the surrounding literature commonly uses the terms interchangeably, and refers to all network-based statistical dependence representations as graphical models.} Unlike broader philosophical notions of causality, Granger causality is a specific statistical concept, where past values of a time series improve the predictability of another beyond what can be achieved using only the target’s own history \cite{Granger,luet}.

While many studies build Granger-causal networks from bivariate effects, the more comprehensive multivariate approach is derived from the popular Vector Autoregression (VAR) framework \cite{luet}. Significant research has extended this baseline, focusing on robust estimation techniques \cite{robust_var_1,robust_var_2,robust_var_application,conf,rank2} and broadening identified effects to include volatility and skew-type relationships \cite{cornell2024enhancingcausaldiscoveryfinancial}.

In this study, we pursue the recently highlighted direction of fitting acyclic processes. DAG-based causal structures have become commonplace following the introduction of Bayesian Networks \cite{pearl1985bayesian}, and their later widespread adoption as a framework for causal analysis \cite{pearl1,pearl2}. Unsurprisingly, this has spurred efforts towards DAG-driven analysis for VAR models. However, they near invariably apply the PC algorithm \cite{PCbook} for Bayesian network identification to the residuals of a standard VAR \cite{PC1, XU2023100229, Awokuse01052003, babula2004modeling}, developing an acyclic contemporary causal network, rather than acyclic Granger causal networks. 

Here, we demonstrate an evolutionary approach to identifying acyclic Granger causal networks through the use of DAG-constrained VAR models. EA's have similarly been applied to Bayesian network identification in simultaneous variable models. These approaches generally use a genetic representation that encapsulates both structural elements of the graph, and exact link details. Common structures include multi lists of topological orderings and associated links \cite{Dai2020, classifier1, classifier2}, or directly utilizing the network adjacency matrix \cite{evoMCMC}. Many studies often involve hybrid techniques that insert guided elements into the formerly random reproduction, aiming to increase search efficiency \cite{carlos2002, evoMCMC, classifier2, Muruzábal2007}. For a detailed discussion of these hybrid approaches see \cite{Muruzábal2007}. Rather than focusing on adjusting reproduction and mutation to improve efficiency, our study introduces an improved representation that inherently reduces the search space by effectively leveraging the computational efficiency of the VAR model.

\section{HEAVe Methodology}

\label{sec:methods}

In this section, we present the methodology for constructing and representing causal networks, focusing on the identification of individual links using a VAR model. We also provide the mathematical framework for DAG-constrained VAR and describe how HEAVe solves this optimization problem.

\subsection{Graphical Structures}
\label{sec:graphical_structures}

A directed graph (digraph) $G$, is traditionally represented as a tuple $(V,E)$ of a set of vertices $V$, and a set of edges $E$. These are often represented together as an $N\times N$ adjacency matrix $W$, where $N=|V|$, with elements $W_{ij}=1$ if there is an edge linking the $i$ and $j$ nodes, and $W_{ij}=0$ otherwise. A Directed Acyclic Graph (DAG) is a specific type of digraph which lacks any cycles, or equivalently, a graph that can be topologically ordered into a list, $O$, of nodes, such that for all edges $e_{ij}\in E$, $i$ comes before $j$ in $O$. An ordering that meets these edge criteria we call $G$ consistent, and there is at least one, and often many consistent orderings for each DAG. Likewise, there are many graphs consistent with a typical ordering.

While orderings provide some information around the relative position of nodes within a DAG, our intuition is often more suited to considering hierarchical representations. Here, we define a graph hierarchy, $H$, to be a list of length $|V|$ where the $j$'th element denotes the hierarchy of node $j$. An edge $e_{ij} \in E$ is $H$ consistent if the edge source, $i$, has a higher hierarchical level than the target, i.e. $H(i)>H(j)$. A graph $G$ is $H$ consistent if all edges $e_{ij} \in E$ are $H$ consistent. For each graph $G$, there exists a set of consistent hierarchies, and if this set is not empty, $G$ is a DAG. Likewise, for a specific hierarchy $H$ there exists an (often large) set of consistent DAGs. 

For a given hierarchy $H$ we can create a matrix of all consistent edges:
\[C_H=\mathds{1}_{H(i)>H(j)}.\]
The set of graphs whose adjacency matrices can be achieved by purely removing elements from $C_H$ are those that are $H$ consistent. This matrix $C_H$ therefore determines the set of consistent graphs, and we henceforth call it the $H$ constraint matrix. We can likewise create an equivalent constraint matrix for the $O$ representation, by having $C_{ij}$ entries compare the relative order of the $i$ and $j$ nodes.

\begin{figure}[H] 
    \centering
    \includegraphics[width=\linewidth]{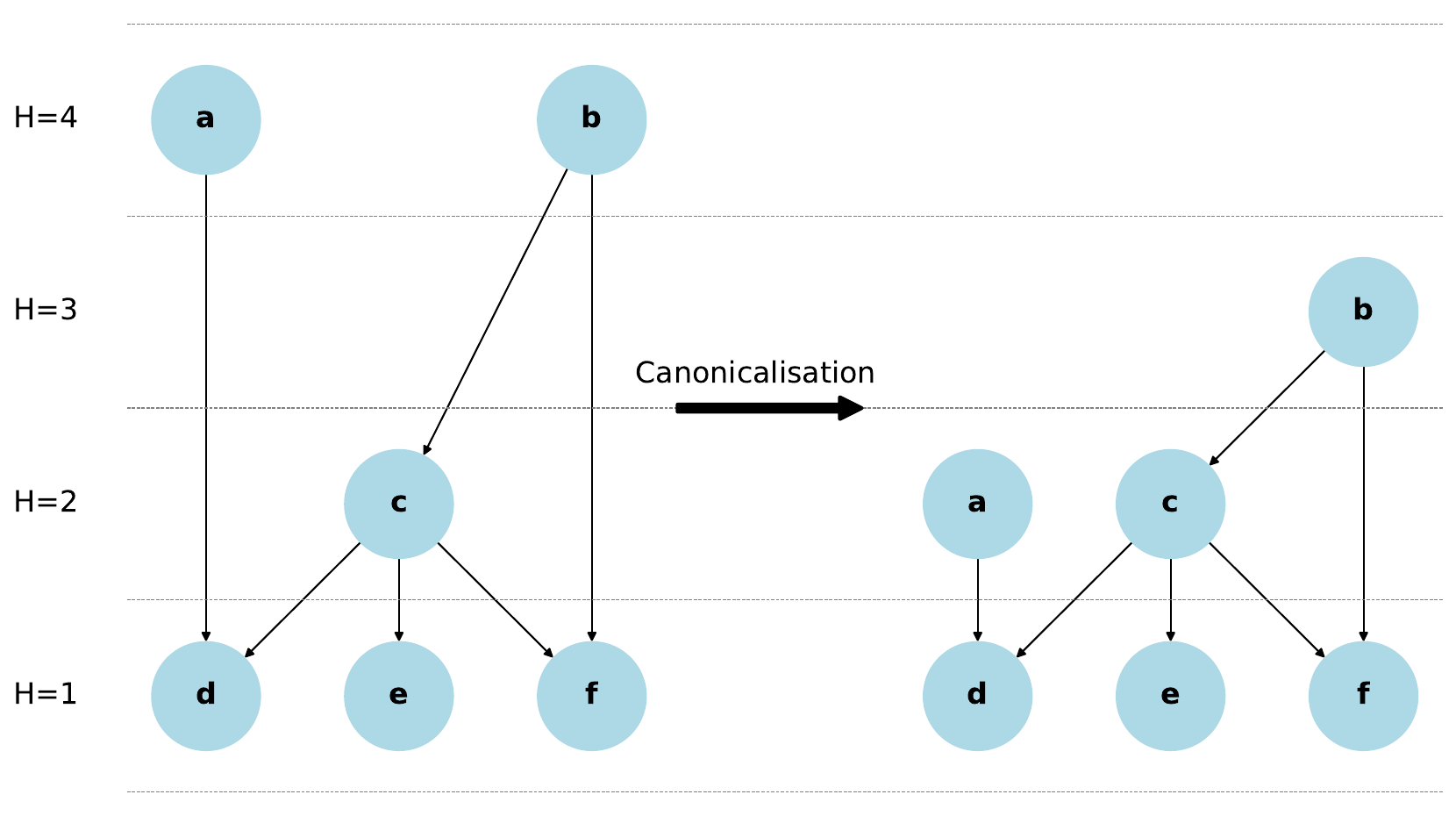} 
    \caption{Canonicalisation of a hierarchy drops nodes to their lowest possible position.}
    \vspace{-2mm}

    \label{fig:canonicalisation}
\end{figure}

Ideally, we would have a single unique hierarchy associated with each DAG, and hence define the canonical hierarchy $H_c$, which is the lowest, positive integer valued consistent list that is consistent with $G$. Algorithmically, this can be calculated from a DAG $G$ by setting all nodes with 0 outgoing edges to $H=1$, and then iterating through a predecessor tree. Notably, this works on any DAG, and does not require it to originate from a hierarchical representation. \autoref{fig:canonicalisation} gives a visualisation of this process.

\subsection{Evolutionary Algorithm}

\subsubsection{Representation}
\label{sec:rep}
Our novel approach directly utilizes the hierarchical list, $H$, as the unit of evolution (candidate solution), $X$, rather than evolving specific DAG instances. This means our EA searches a reduced, hierarchical space, where individual solutions/members of our population, $X_i$ are associated with multiple DAG candidates. The benefits of this approach are twofold: 

\begin{enumerate}
    \item Separating link and structure identification shrinks the search space for the evolutionary process by efficiently avoiding cyclic graph structures. Further, the inherently acyclic nature of the representation ensures the preservation of acyclicity during reproduction and mutation.
    \item Using the hierarchical list as the unit of evolution provides a relatively smooth evolutionary surface, which is especially advantageous when the analysis of this hierarchy is our ultimate goal.
\end{enumerate}

One critical consideration when implementing a hierarchical algorithm is the treatment of boundaries. While it may seem natural to bound $H$ values between $1$ and $N$, for a network of $N$ nodes, it is important to note that the $H$ values are only relevant for constructing the constraint matrix $C$, which depends solely on the relative rather than absolute values of $H$ across nodes. Moreover, the canonical hierarchy $H_C$, used when analyzing the developed DAG, is naturally bounded, rendering explicit bounds during runtime unnecessary. Enforcing such bounds may also hinder the algorithm's ability to resolve relative positions near the boundaries. For example, if all nodes are at $H=1$, but one node should actually remain at $H=1$ while the others move to $H=2$, a bounded approach would require shifting all nodes upward, whereas an unbounded approach could simply drop the single node down.

Another consideration is the treatment of un-utilized hierarchical layers. For instance, if nodes occupy $H=1,2$, and $4$ but not $3$ (as we saw in in \autoref{fig:canonicalisation}), should the intermediate layer be removed? We argue that these intermediate layers provide value in resolving causal relationships. For example, if nodes $i$ and $j$ both occupy $H=2$ and there exists a true link $e_{ij}$, there will be downward pressure on $j$. However, $j$ might also have links to nodes at $H=1$, making it difficult to resolve without shifting entire layers. Allowing these un-utilized layers aids the algorithm in resolving local relationships without disrupting the global hierarchy.

Both removing bounds and allowing for un-utilized layers improve the navigability of the optimization space (akin to convexity in continuous functions) by reducing local minima. These design choices create a landscape that supports the local greedy adjustments made by an EA as valid and effective paths toward the optimal hierarchical structure.

To assess these ideas, in \autoref{sec:sim_study}, we evaluate three variants of $H$ algorithms: one unbounded (`floating'), one bounded, and one that re-canonicalizes the hierarchy at each iteration by removing unused layers.

\subsubsection{Parent selection}
Selective reproduction serves as the primary exploitative element of evolutionary optimization, pushing subsequent generations to score better on our objective function. This process plays out through two elements: the allocation of reproductive fitness across parents, and the generational mechanics of the population. 

Aside from the selection of parents, at the population level reproduction is generally implemented through either a fixed or variable population size algorithm. In the fixed size variant, each generation we create a predetermined number of new members with parents selected according to some reproductive fitness measure, $R(X_i)$. For the flexible size algorithms, each member $X_i$ has a chance to produce a variable number of children according to its reproductive fitness, with the total number of children unconstrained. Each of these approaches has advantages and disadvantages, however for our purposes of algorithmic comparison the fixed population approach has fewer stochastic elements, and produces more statistically regular outcomes. For similar stability/simplicity reasons, our implementation has complete re-population at each generation, with none of the previous parents surviving. This is commonly known as $(\lambda, \mu)=(N,N)$ reproduction. The parents of each of the $N$ new members is determined by randomly selecting according to the probabilities: 
\begin{align}
  P(X_i)=\frac{R(X_i)}{ \sum_j R(X_j)}.  
\end{align}


A simple way to score the reproductive success of each member would be to linearly scale $R(X_i)$ to the objective function: $R(X_i)=f(X_i)$.
The immediate limitation to this approach is the non-standardized variability in $f(X_i)$. For some problems, good solutions may score several magnitudes higher than others, but we may encounter problems where a good solution is only very marginally better than a poor solution, and the reproductive pressure is insufficient. We could center and standardize all $f(X_i)$, however the distribution of these values will still be \emph{a priori} unknown. Instead, we utilize rank-based reproductive fitness, such that the $Rank(f(X_i))$ will always be scattered in a pre-determined distribution. Here we have:
\begin{align}
    R(X_i)=|\{X_j: f(X_j) < f(X_i)\}|+0.5|\{X_j: f(X_j) = f(X_i)\}|+1.
\end{align}

\subsubsection{Recombination}
After selection, the parents must undergo a recombination method to determine the genetic information of their offspring, $X'$. For our hierarchical representation this is a relatively simple process, where we randomly select each of the $X_j$ elements to be from either parent $A$ or $B$:
\begin{align}
  X'_{A,B}=IX_A+(1-I)X_B,  
\end{align}
where \( I \sim \text{Bernoulli}\left(0.5\right) \). The recombination of a permutation is comparatively more complex. We begin by randomly selecting half the entries of parent A and copying them into the blank child: 
\begin{align}
X'_{A,B}=IX_A,    
\end{align}
with $I$ as before. We then iteratively populate the unused positions in $X'$ with the highest order yet-unused elements of parent $B$. This could be considered a variation to the common order crossover for permutations \cite{into_to_EC}, where instead of one large block with remainder fill-in, we take many smaller blocks.

\subsubsection{Mutation}

After a reproductive step evolutionary algorithms generally look to mutate the resulting individual. This serves as the primary means of exploring new regions in the state space, and preserving genetic diversity. Starting from the new individuals hierarchy list, $X'$,  for all nodes with some probabiltiy $p$ we mutate that nodes $H$ score up or down according to some mutation density:
\begin{align}
    X'=X+I \cdot M,
\end{align}

\noindent where the mutation indicator variable \(I\) is defined as:
\begin{align}
I_i =
\begin{cases} 
1, & \text{with probability } p/2, \\
0, & \text{with probability } 1 - p, \\
-1, & \text{with probability } p/2,
\end{cases}
\end{align}
and $M$ is some positive, interger valued step size density. We would like this density to be strictly decreasing, such that smaller mutations are always more likely than large ones. This, combined with the interger requirements suggests a natural choice is the geometric distribution with some mean step size parameter $\lambda$ ($\lambda=1/p$ in terms of the standard parametrisation).

We apply a similar geometrically distributed mutation to orderings, now shifting positions in the string, rather than scores: 
\begin{align}
X' = \Pi(X, I \cdot M),    
\end{align}

\noindent where \( \Pi(X, S) \) represents the permutation of \( X \) based on the mutation vector \( I \cdot M \), where elements $X_i$ are shifted left or right based on $(I \cdot M)_i $. This shifting is done left to right (highest position nodes first), and capped at the vector edges (no cyclic shifting), an approach often known as swap mutation for permutations \cite{into_to_EC}.

\subsection{Vector Autoregression}
\label{sec:VAR}
Vector autoregression (VAR)  \cite{sims_1980} models the joint dynamics and causal relations among a collection of time series. It is the natural multivariate extension of the univariate autoregression (AR) model frequently used to analyse the inter-temporal dependency of a sequence of observations. Under the VAR(p) formulation the expectation of the data vector $\boldsymbol{y}_t$ at the next observation is a linear function of $p$ previous observations. Equations 1 and 2 below show the relationship for order-1 and order-p lagged variants: 
\begin{align}
\mbox{Order-1:} \;\;\; & \boldsymbol{y}_t =A_1\boldsymbol{y}_{t-1}+\boldsymbol{c}+\boldsymbol{\epsilon}_t , \label{eq:var_lin} \\
\mbox{Order-$p$:} \;\;\; & \boldsymbol{y}_t =A_1\boldsymbol{y}_{t-1}+A_2\boldsymbol{y}_{t-2}+...+A_{t-p}\boldsymbol{y}_{t-p}+\boldsymbol{c}+\boldsymbol{ \epsilon }_t , 
\end{align}

\noindent where $\boldsymbol{y}_t \in \mathbb{R}^N$ contains observations at time $t$, $\boldsymbol{c} \in \mathbb{R}^N$ is constant, the $A_k \in \mathbb{R}^{N\times N}$ are coefficients for lags $k=1,...,p$, and $\boldsymbol{\epsilon}_t \in \mathbb{R}^N$ contains error terms with zero mean and covariance matrix $\Sigma_\epsilon$. The $\boldsymbol{\epsilon}_t$  is often assumed to be Gaussian, however this is not a strict requirement. The VAR model assumes the current value of each variable depends on its past values and those of potentially all other variables in the system (full conditioning).

Estimating a VAR model involves determining the coefficient matrices $A_k$ and the error covariance matrix $\Sigma_\epsilon$. Since our focus is on the causal influence structure within the dataset, our primary interest lies in estimating $A_k$, as these matrices fully define the causal relationships and, consequently, the desired Granger causal network $G$. This estimation is typically achieved using the multivariate least squares (MLS) method, which frames the problem as a general multivariate regression:
\begin{align} \min_A |AY_{t-1} - Y_{t}|, \end{align}
where $Y_t=[\boldsymbol{y}_1, ...,\boldsymbol{y}_T]$. The MLS method provides a closed-form solution via orthogonal projection \cite{luet}.

The structure of the Granger causal network $G$ is determined by the statistically significant elements of $A^T_k$. Under finite variance assumptions, the estimates $A_k$ are asymptotically normally distributed:
\begin{align}
    \sqrt{N} \; \mbox{Vec}(\hat{A}_k-A_k)\xrightarrow[]{d}\mathcal{N}(0, \Gamma^{-1}\otimes\Sigma_\epsilon) , \label{eq:kron}
\end{align}
where $\Gamma=\ YY'/N$, $\otimes$ indicates the Kronecker product and $\mbox{Vec}(\cdot)$ denotes casting a matrix into vector form. For the case of a VAR(1) model the term $Y$ is the matrix representation of our response data $\boldsymbol{y}_t$, implying that $\Gamma$ is an estimate of the covariance matrix of returns. For generalised VAR(p) the complexity of this matrix increases, however Equation \ref{eq:kron} is still valid. 

To evaluate the presence of a link $e_{ij}$, we construct $t$ values associated with the null hypothesis $A_{i,j}=0$ as $t_{ij}=\hat{A}_{i,j}/\hat{s}_{i,j}$, where $\hat{s}_{i,j}$ represents the corresponding standard error, derived from $\Gamma^{-1}\otimes\Sigma_\epsilon$. Links in $G$ are identified with a false positive probability $\alpha$ as $e_{ij}=|t_{ij}|>\Phi(1-\alpha/2)$, where $\Phi$ is the inverse normal CDF.

To streamline our discussion and estimation of causal networks, we limit our analysis to VAR(1) processes and omit the index $p$ from our discussion, with $A=A_1$ unless otherwise specified. 
For the empirical networks in Section \ref{sec:emp_results} we operate under the assumption that any causal link $i\rightarrow j$  will first manifest in order-1 effects, and that lag $p>1$ effects will not occur independently of a 
$p=1$ dependence. This assumption is intuitive and the scenarios where it doesn't hold are expected to be relatively rare.

In total, the estimated Granger causal network $\hat{G}$ is derived as: 
\begin{align}
    \hat{G} = \mathbb{I} \left (\frac{|\hat{A}|}{S} > \Phi(1 - \alpha/2)\right), 
\end{align}
where $S$ is the matrix of standard errors. This relationship is crucial to note, as while we have primarily discussed concepts in the binary matrix $G$ space, the underlying mathematical framework resides in the continuous-valued space of recurrence matrices, $A$.  

\subsection{Acyclic VAR and evolutionary objective function}

Working directly in the space of VAR recurrence matrices, $A$, the acyclic constraints would enter the default VAR objective function (left term below) in the form of complex matrix polynomial conditions:
\begin{align} \min_A |AY_{t-1} - Y_{t}| \quad \text{s.t.} \quad \text{Diag} (A^n) = \mathbf{0} \, \forall \, n > 1, \label{eq:default_opt} \end{align}
where $Y_t=[\boldsymbol{y}_1, ...,\boldsymbol{y}_T]$. 
While the function we're minimizing remains simple, ensuring candidate matrices continue to satisfy the constraints is quite difficult, and even verifying whether a solution satisfies the conditions requires nontrivial computation.

\begin{figure}[h!] 
    \centering
    \includegraphics[width=\linewidth]{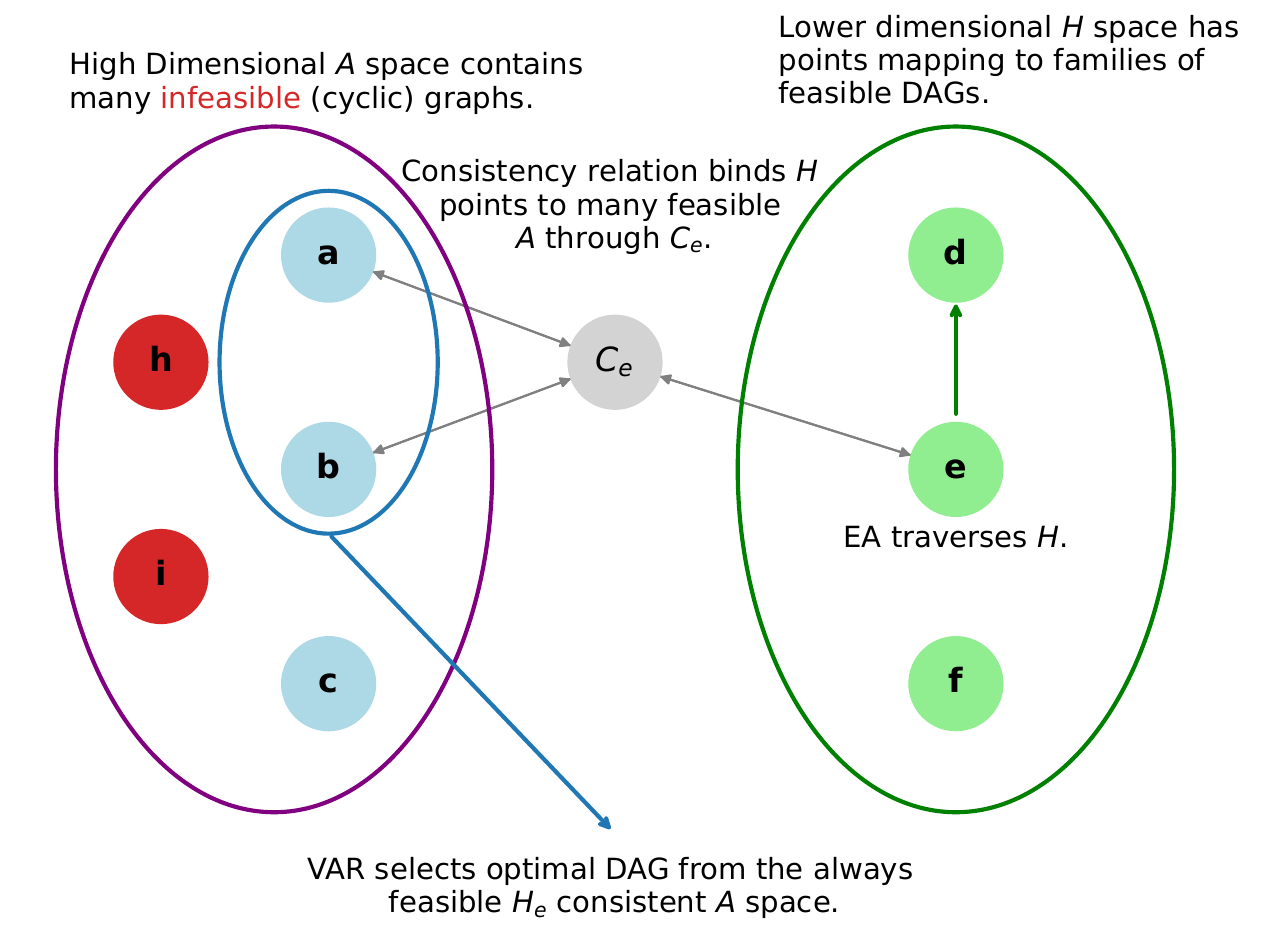} 
    \caption{Illustration of HEAVe’s indirect traversal of A-space through H-space EA and local VAR optimizations.}
    \vspace{-2mm}
    \label{fig:Diagram}
\end{figure}

\noindent Our proposed solution decouples the problem into two stages, with the EA operating across a reduced space and VAR performing local optimizations. \autoref{fig:Diagram} gives a visualisation of this process. From the perspective of the EA, each proposed solution, $X_i$ (hierarchy list or node ordering) generates an associated constraint matrix $C_{X_i}$, which defines the permissible links in $A$. Identifying the optimal $C_{X_i}$ consistent recurrence matrix $A$ can then be formulated as:
\begin{align} \min_A |AY_{t-1} - Y_{t}| \quad \text{s.t.} \quad (\mathbf{1} - C_{X_i}) \circ A = \mathbf{0}, \label{eq:local_opt} \end{align}


where $\circ$ denotes the element-wise (Hadamard) product. Fortunately, multi-target linear regression can be decomposed into a series of parallel, independent regressions. On this per-target basis the constraints can be trivially satisfied by including only those covariates with permissible links to the target variable, as defined by $C_{X_i}$.

For our EA, Equation \ref{eq:local_opt} serves as the objective function $f(X_i)$, which evaluates the success of each evolutionary unit. The EA operates by searching the space of constraint matrices, where each solution proposes a set of consistent graphs, which the VAR framework then refines into a specific estimated DAG. The complete optimization process can be expressed as:
\begin{align} \min_X \left( \min_A |AY_{t-1} - Y_{t}| \quad \text{s.t.} \quad (\mathbf{1} - C_{X_i}) \circ A = \mathbf{0} \right), \label{eq:complete_opt} \end{align}
where the inner convex and continuous problem is efficiently solved using least squares, while the outer discrete problem is addressed by the EA.

It is very important to also note that all $C_{ii}=1$, i.e. we allow self loops during runtime. Technically speaking, this violate the notion of a DAG. However, empirically, self causal effects are far too common to ignore. For DAG loyalists, we can simply ignore the diagonals, treating them as control variables when analyzing results, or not-allow self-loops, but pre-filter the data to be the residuals of univariate auto-regressions (letting self-causal effects a priori dominate cross-causal effects is not without merit).

\subsection{Metrics}

One key motivation for fitting causal DAGs to time series data is to uncover the potential hierarchical structure of the underlying system. Beyond evaluating predictive power—which depends on the fitted recurrence matrix , $\hat{A}$, we assess the quality of the hierarchical structure derived from the fitted DAG. Specifically, we focus on the canonical $H_c$ obtained from $\hat{G}$, as described in \autoref{sec:graphical_structures}. To quantify the accuracy of this inferred hierarchical ordering, we propose the hierarchical score ($H$ score), which evaluates the concept of permissible links.

For each potential link $\hat{e}_{ij}$ in the fitted network $\hat{G}$, we determine whether the link satisfies the condition $H(i) > H(j)$ under the associated canonical hierarchy $\hat{H}_C$ and compare this with a ground truth hierarchy $H_C$ derived from the true causal DAG. The hierarchical score is calculated as the proportion of links for which these labels align. It is important to note that this metric is only applicable in simulations, where the true hierarchical structure is known, and cannot be directly incorporated into the training process for empirical datasets.

The objective function, by contrast, primarily evaluates the predictive quality of links, including those ultimately deemed statistically insignificant. While the hierarchical score addresses structural ordering between nodes, these metrics individually lack a comprehensive evaluation of the complete binary causal DAG $\hat{G}$. To address this, we introduce the F1 classification score, which evaluates $\hat{G}$ against the true structure $G$. This score, defined as the harmonic mean of precision and recall, is computed by comparing all binary links $\hat{e}_{ij} \in \hat{G}$ to their true counterparts $e_{ij} \in G$.

Lastly, when evaluating the objective function, $f$, we choose to standardize the raw value against the unconstrained VAR model, $f^*=f_{DAG}/f_{VAR}$. This controls for the variability in the per-graph predictability, which significantly increases the statistical stability when comparing algorithms. As the default VAR provides the best in sample linear fit, this functions effectively as an upper bound to standardize with.

\section{Simulation Study}
\label{sec:sim_study}

As discussed in \autoref{sec:methods}, our goal is to generate a directed acyclic graph $G$ that captures the causal relationships within a system of time series, enabling us to infer hierarchical structures. In practice, real-world datasets typically lack a "ground truth" graph and  associated hierarchy against which our inferred results can be validated. As a result, this task becomes analogous to unsupervised learning of $G$ and $H$, where the underlying structures must be inferred without direct supervision. This limitation makes it impossible to evaluate our methodology’s ability to identify $G$ and uncover hierarchies using empirical datasets alone. Instead, we rely on simulations where the true underlying structure is explicitly controlled, allowing us to evaluate the accuracy and robustness of our approach. Additionally, simulations provide the advantage of generating numerous graphs and datasets while allowing precise control over the structural parameters of the graph, such as sparsity, hierarchy depth, and edge density. This section outlines the methodology used to simulate DAG-VAR processes, the theoretical underpinnings of the simulation design, and an analysis of our method’s performance across a range of controlled datasets.

\subsection{Simulating VAR processes}

A VAR process is described by the constant term $\boldsymbol{c}$, the recurrence matrices $A_i$, and the error process distribution, $F_\mathcal{E}(\boldsymbol{\epsilon}_t)$, which is designed to have an expected value of $0$. A typical simplifying assumption is that $\boldsymbol{\epsilon}_t$ follows a Gaussian distribution, making the entire process fully characterized by the covariance matrix $\Sigma\epsilon$.

\subsubsection{Generating $\Sigma_\epsilon$:}
\label{sec:generating_sigma}

The first step involves generating a random covariance matrix that satisfies the conditions of symmetry and positive definiteness. This is accomplished by first sampling a set of positive eigenvalues, ${\lambda_i}$, as the absolute values of standard normal random variables: ${\lambda_i} \sim |\mathcal{N}(0,1)|$. A random rotation matrix is then derived by extracting the Q term from the QR decomposition of a randomly generated matrix with entries drawn from a standard normal distribution. The resulting covariance matrix is then: \begin{align} \Sigma_\epsilon = Q D(\lambda_i) Q^T, \end{align} where $D(\lambda_i)$ represents the diagonal matrix of the eigenvalues.

\subsubsection{Generating A:}
\label{sec:gen_a}

The second step involves constructing valid coefficient matrices that ensure the stationarity of the process. To achieve this, the eigenvalues of $A$ must reside strictly within the unit circle. Eigenvalues on the perimeter ($|\lambda_i| = 1$) would result in a drifting process, while eigenvalues with $|\lambda_i| > 1$ would produce a divergent process. Using the same method employed to generate $\Sigma_\epsilon$ would yield dense coefficient matrices, which are unsuitable for approximating our intended causal structure.

Instead, we generate a sparse Erd\H{o}s-R\'{e}nyi graph, where each edge is selected with some probability $p$, and assign Gaussian-distributed values to the edges to construct $A^*$. The resulting matrix is normalized by dividing by approximately the absolute value of its largest eigenvalue: $A={A^*}/{\max(|\lambda_i|\times1.05)}$ where the factor $1.05$ ensures sufficient distance from the boundary of stationarity to avoid drift.

\subsubsection{DAG enforcement:}
To ensure that our randomly generated VAR recurrence matrix, $A$, adheres to DAG constraints we begin by generating a random hierarchical list, $X \sim U[1, N]^n$, equivalent to the initialization process for the evolutionary algorithm. We then construct the constraint matrix $C_X$ associated with $X$. The Hadamard product $A\circ C_X$ will then be a DAG consistent recurrence matrix. Notably, this process removes approximately 50\% of the original links (formally, (N+1)/2N\%). Consequently, for our simulations, the effective link rate is half the initial rate $p$.

\subsection{Parameters}

We generate 30 synthetic datasets along with ground truth causal networks using the following parameters:

\begin{itemize}
    \item[\textbullet] $N=30,60,100$ nodes per process (corresponding to 900, 3,600, and 10,000 potential links) with $30\times N$ observations per network (rows in the data matrix).
    \item[\textbullet] Recurrence matrix and Covariance matrix designed according to \autoref{sec:gen_a} and \autoref{sec:generating_sigma}, with a constant $c=0$.
    \item[\textbullet] Proportion of links $p=0.25$, effectively reduced to $0.125$ after DAG enforcement.
\end{itemize}

We apply the four methods described in \autoref{sec:rep} to these datasets, with the following parameters:

\begin{itemize}
    \item[\textbullet]  A 10\% mutation chance, with an expected step size of $N/10$ (1\% effective mutation rate).
    \item[\textbullet] Population size of 30 members (increasing this parameter showed minimal benefit relative to the computational cost).
\end{itemize}

It is important to note that these parameters have not been tuned to maximize performance, but rather are chosen as a simple baseline for algorithmic validation and comparison.

\subsection{Simulation Results}
\begin{table}[h!]
    \centering
    \vspace{-3mm}
    \caption{Summary statistics for simulation results, calculated over the final 5 generations of each simulated graph. Displaying the mean of the following statistics: the population average (standardized) objective function score: $f^*$, the objective function score for the best-performing member of the population: $\hat{f^*}$, the population average hierarchical score $H$, and the population average $F1$ classification score.}
    \vspace{-2mm}
    \label{tab:model_comparison}

    \begin{subtable}{\columnwidth}
        \centering
        \caption{30 Nodes}
        \label{tab:30nodes}
        \begin{tabular}{l S[table-format=2.4] S[table-format=2.4] S[table-format=2.4] S[table-format=2.4] }
            \toprule
            \textbf{Model} & \textbf{$f^*$} & \textbf{$\hat{f^*}$} & \textbf{$H$} & \textbf{$F1$} \\
            \midrule
            Floating & 0.927 & 0.935 & 0.847 &  0.758 \\
            Canonical & 0.806 & 0.865 & 0.735 &  0.577\\
            Bounded & 0.909 & 0.929 & 0.844 &  0.733\\
            Ordered & 0.878 & 0.914 & 0.798 &  0.670\\
            \bottomrule
        \end{tabular}
    \end{subtable}

    \vspace{0.15cm}

    \begin{subtable}{\columnwidth}
        \centering
        \caption{60 Nodes}
        \label{tab:60nodes}
        \begin{tabular}{l S[table-format=2.4] S[table-format=2.4] S[table-format=2.4] S[table-format=2.4] }
            \toprule
            \textbf{Model} & \textbf{$f^*$} & \textbf{$\hat{f^*}$} & \textbf{$H$} & \textbf{$F1$} \\
            \midrule
            Floating & 0.919 & 0.929 & 0.845 &  0.714 \\
            Ordered & 0.853 & 0.883 & 0.767 &  0.601\\
            \bottomrule
        \end{tabular}
    \end{subtable}

    \vspace{0.15cm}

    \begin{subtable}{\columnwidth}
        \centering
        \caption{100 Nodes}
        \label{tab:100nodes}
        \begin{tabular}{l S[table-format=2.4] S[table-format=2.4] S[table-format=2.4] S[table-format=2.4] }
            \toprule
            \textbf{Model} & \textbf{$f^*$} & \textbf{$\hat{f^*}$} & \textbf{$H$} & \textbf{$F1$} \\
            \midrule
            Floating & 0.916 & 0.924 & 0.856 &  0.695 \\
            Ordered & 0.846 & 0.871 & 0.750 &  0.584\\
            \bottomrule
        \end{tabular}
    \end{subtable}


\end{table}

\subsubsection{Comparison of Hierarchical algorithm variants}

\begin{figure}[H]
    \centering
    \includegraphics[width=\columnwidth]{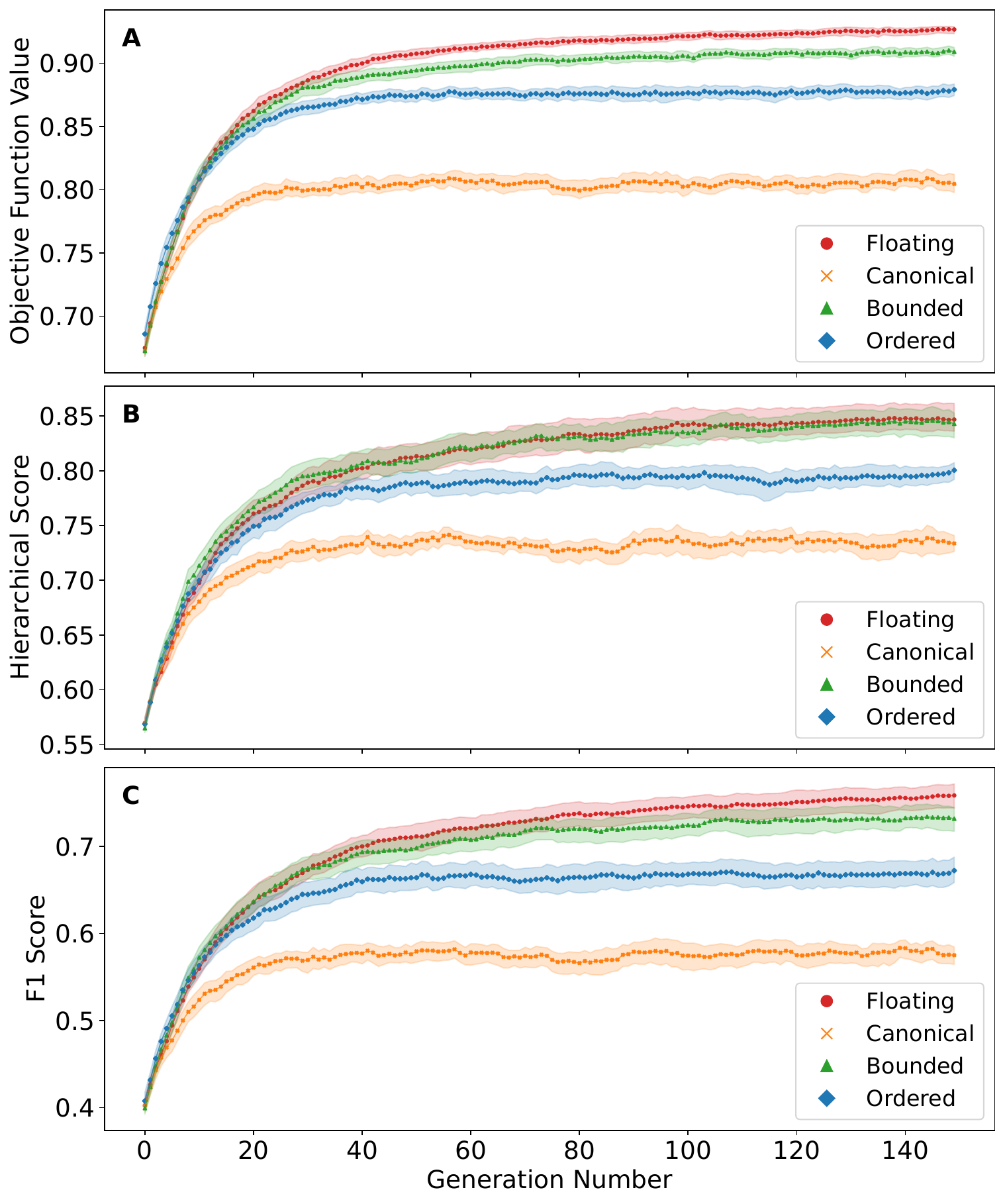} 
    \caption{Performance metrics per generation, showing both the sample mean and bootstrapped 95\% confidence intervals for this mean across the several hierarchical representations as well as our order-based comparison. The three subplots represent: (a) Mean objective function value, (b) Mean hierarchical score, and (c) Mean F1 link classification score.}
    \vspace{-2mm}
    \label{fig:H_variants}
\end{figure}

\autoref{fig:H_variants} displays several statistics on a per generation basis for the $N=30$ networks, alongside bootstrapped 95\% confidence intervals for these values. Observing this Figure in conjunction with the results from \autoref{tab:30nodes}, which documents the means of these values across the final 5 generations we draw the following conclusions: 


\begin{enumerate}
    \item Free floating optimization performs either equally well, or superior to the bounded H representation. It achieves better objective function score (0.927 vs 0.909), slightly better F1 (0.758 vs 0.733) score and roughly equal H score.
    \item Canonicalisation  consistently reduces performance across all three metrics, with less smooth convergence consistent with our notion of the non-convexity of the constraint effects. 
    \item Both non-canonical forms of hierarchical representation perform substantially better than the ordering based algorithm, with the largest difference being a 13.1\% higher F1 classification score for the floating H versus the ordering algorithm. 
\end{enumerate}

\autoref{tab:60nodes} and \autoref{tab:100nodes} provide some insight into the scaling of our algorithm against network size. \autoref{tab:60nodes} shows results of our best performing H-algorithm (free floating) against the ordering representation for $N=60$. The results are consistent with \autoref{tab:30nodes}, with very similar scores across the board. The H-algorithm achieves functionally equivalent $f^*$ and $H$ scores, with slightly reduced $F1$ score, while the ordered algorithm has slightly depressed scores across all metrics. We consider even larger networks in \autoref{tab:100nodes} and see a near identical pattern. The \% improvement of the hierarchical algorithm compared to the order based approach is increasing in network size, primarily through a functional absence of the $f^*$ and $H$ score performance decay that the order based approach suffers, and a significant reduction in the rate of $F1$ score loss. 

\section{Empirical study}
\label{sec:emp_results}
\subsection{Data}

Our dataset consists of hourly price data for 100 cryptocurrencies spanning the period from January 1, 2021, to January 1, 2022, along with their market capitalizations as of June 2022 \cite{cornell2023vector}. Market capitalization is calculated as the product of a cryptocurrency's price and its circulating supply. The cryptocurrencies were selected based on their market capitalization rankings as of June 2022, though some deviations from the exact top 100 occurred due to incomplete or missing price data. Hourly returns, denoted by $y_t$, were computed as the natural logarithm of the ratio between consecutive prices $p_t$ in the Coin/USD exchange rate, expressed as $y_t = \log(p_t/p_{t-1})$.

\subsection{Empirical Graphs}

\begin{figure*}[ht!] 
    \centering
    \begin{subfigure}[t]{0.48\textwidth}
        \centering
        \includegraphics[width=\textwidth]{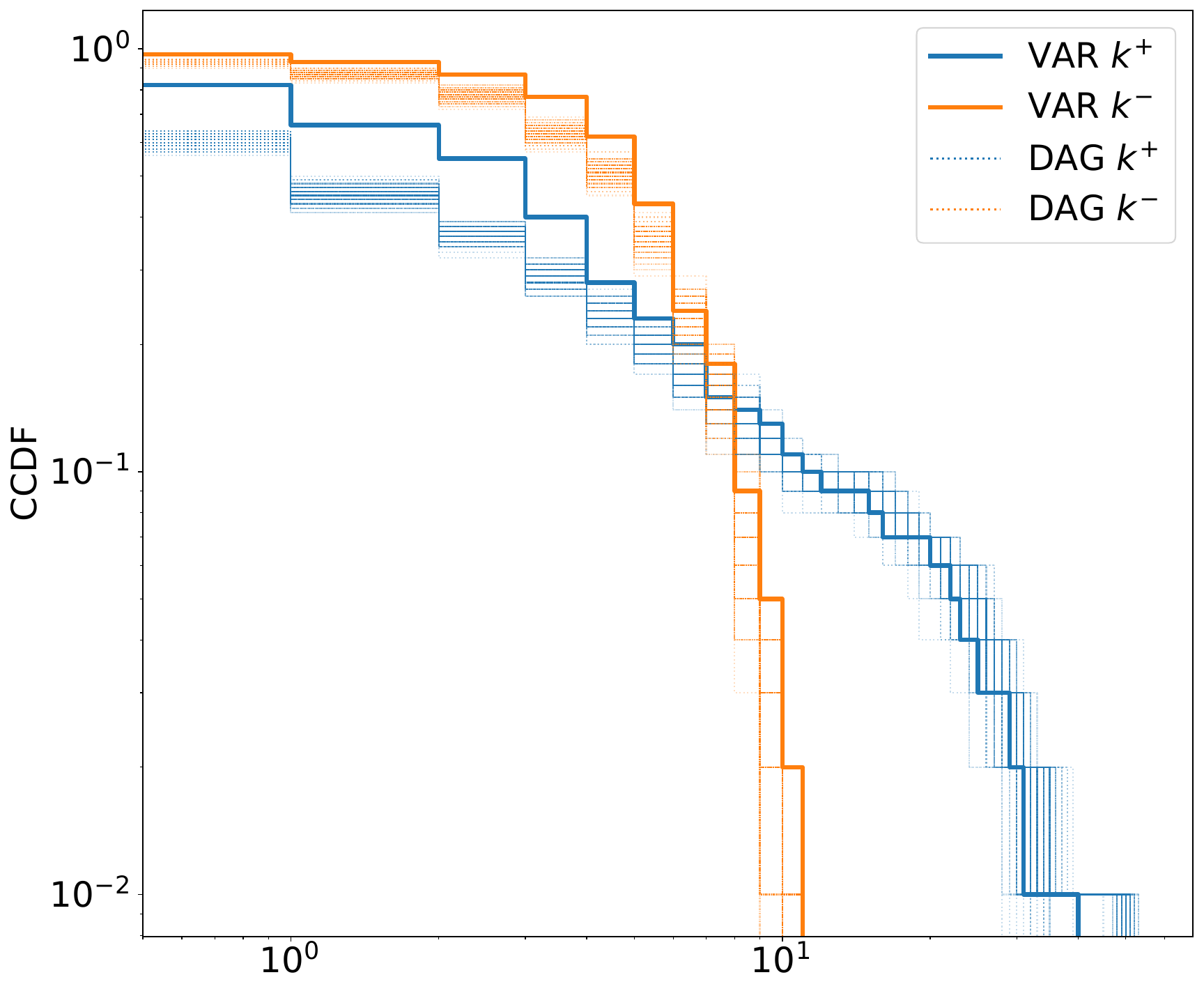}
        \caption{Degree distribution CCDF plots, comparing the unconstrained network against our evolutionary samples.}
        \label{fig:ccdf-plot} 
    \end{subfigure}
    \hfill
    \begin{subfigure}[t]{0.48\textwidth}
        \centering
        \includegraphics[width=\textwidth]{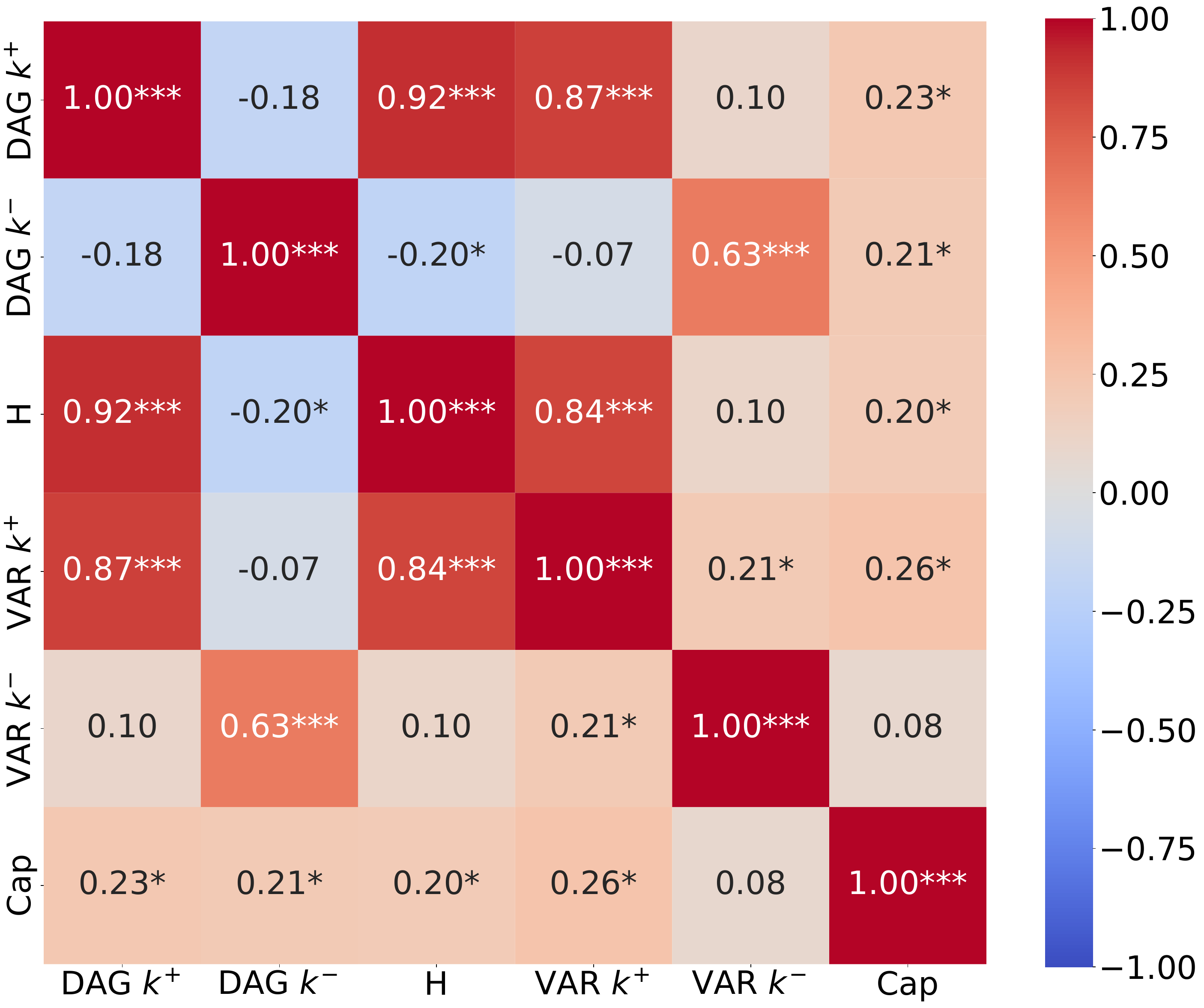}
        \caption{Spearman rank correlation ($\rho$) heatmap of key node attributes, with associated statistical significance levels indicated by asterisks($*\rightarrow p\leq0.05, **\rightarrow p\leq0.01, ***\rightarrow p\leq0.001$).}
        \label{fig:heatmap-plot} 
    \end{subfigure}
    \vspace{-2mm}
    \caption{Empirical Network node attributes.}
    \vspace{-2mm}
    \label{fig:emp_plots} 
\end{figure*}

When analyzing the output of our evolutionary algorithm in detail, we observe an additional benefit: the algorithm produces a collection of graphs rather than a single estimate. This collection allows us to investigate the certainty of a node's hierarchical position through it's empirical variation (however structural correlation will induce a downward bias in $\hat{\sigma}_H$). Further, while we do not formally demonstrate this here, it is intuitive to assume that averaging structural properties across the graphs may provide a more robust estimate than relying on a single graph. However, it is important to note that constructing an `average graph' by simply averaging adjacency matrices would violate the constraints of a directed acyclic graph (DAG) and is therefore not feasible. This highlights the need to carefully interpret the collection of outputs rather than attempting to collapse them into a single structure.

To analyze our results, we examine both the DAG with the highest $f^*$-score (Best DAG), and take averages of several metrics across the graphs generated during the final five iterations (Mean DAG). \autoref{tab:emp_results} summarizes these statistics and compares them to those of the unconstrained model. We first note that the average graph in the terminal samples achieves an $f^*$-score comparable to those observed in our simulated datasets, retaining 92.1\% of the predictive capacity of the unconstrained model. In terms of total links, the terminal sample contains 90.2\% as many edges as the unconstrained model (5.55 vs. 6.15). In converse terms, we have pruned 9.8\% of links and lost 7.9\% accuracy. We record a relatively low average coefficient of variation ($CV=$ standard deviation/mean) for $H$ (0.21), suggesting that generally each nodes $H$ score is relatively stable. At a first glance, we observe a strong positive Spearman correlation ($\rho = 0.51$) between $CV_H$ and the mean $H$ values, however there is a large point mass in the $CV_H$ distribution, with $21/100$ nodes having a constant position, $H=1$, with $CV_H=0$ across all networks. Filtering out this point mass, there is actually a negative correlation ($-0.248$) between position and variation. This pattern is not un-intuitive, as it suggests there is a number of clearly non-influential nodes ($H=1$ constant), a number of clearly high influence nodes, and the greatest variation/uncertainty in the intermediate regions.

\autoref{fig:ccdf-plot} compares the complementary cumulative distribution function (CCDF) of the node degrees from the unconstrained model to the CCDFs of the terminal evolutionary samples
. The evolutionary samples exhibit degree distributions that closely align with those of the unconstrained model, suggesting that this macro-level view of influence dispersion is largely preserved in the constrained models. The observed vertical differences between the VAR line and the evolutionary samples appear concentrated in the lower out degree regions. To test this, for each node we consider the smoothed ratio of out degrees, $k_{DAG}^+ / (k_{VAR}^+ +1)$, and note there is a significant spearman correlation of $0.52$ between this ratio and the associated coins market capitalization. This implies that the majority of the total link count difference (6.15 vs. 5.55) arises from the removal of links associated with lower-influence nodes. This is not entirely unexpected and may suggest that low-degree regions are more influenced by noise, making their links the first to be pruned in order to satisfy the DAG constraints. This effect is also somewhat structurally expected if you take the correlation between hierarchical position and unconstrained out-degree as a given.

\autoref{fig:heatmap-plot} displays a Spearman correlation heat map of node attributes. It includes the correlations of average in-degree, out-degree, and hierarchical position with node degrees in the unconstrained network and market capitalization. As expected from \autoref{fig:ccdf-plot}, node degrees are highly correlated between the constrained and unconstrained scenarios (0.87 $k^+$ and 0.63 $k^-$). Within the acyclic networks, a node's hierarchical position shows a strong correlation with its out-degree (0.92) and a slightly weaker correlation with its unconstrained counterpart (0.84). While this is structurally expected (higher $H$ means more outgoing links are allowed), from such effects we would expect to see an equivalently strong negative correlation for in degree, however this is empirically much weaker (-0.2). Market capitalization exhibits moderate correlations with out-degree (0.23 $k^+$ and 0.21 $k^-$) and hierarchical position (0.20).

\begin{table}[th!]
    \centering
    \caption{Summary of empirical network properties. The table displays the raw objective function score, $f$, the average node degree, $\Bar{k}$, the height of the network hierarchy, and the coefficient of variation for the hierarchical score, $CV_H$.}
    \vspace{-2mm}
    \label{tab:emp_results}
    \begin{tabular}{l S[table-format=2.4] S[table-format=2.4] S[table-format=2.4] S[table-format=2.4] }
        \toprule
         \textbf{Model} & f & $\Bar{k}$ & \textbf{Height} & \textbf{$CV_H$} \\
        \midrule
        VAR & 0.194 & 6.15 & \text{-} & \text{-} \\
        Mean DAG & 0.175 & 5.55 & 14.2 & 0.211 \\
        Best DAG & 0.175 & 5.56 & 14 & \text{-} \\
        \bottomrule
    \end{tabular}
\end{table}
\vspace{-2mm}

\section{Conclusion}
This study introduces a novel approach to fitting acyclic Granger causal networks for time series systems. Our hierarchical evolutionary algorithm, HEAVe, is validated on synthetic networks of varying sizes, where it retains a larger percentage of the unconstrained model’s predictive capacity compared to permutation-based EAs. This advantage extends to network structure discovery, achieving higher link precision and more accurate hierarchy detection. In an empirical study of 100 cryptocurrencies, the identified acyclic networks retain 92\% of the predictive accuracy of unconstrained models while containing 90.2\% as many links. Our results suggest that these networks predominantly function as sub-graphs of the unconstrained networks, consistent with the underlying causal dynamics being hierarchical in nature. This study establishes a baseline for applying EAs to acyclic Granger causal networks and introduces a novel hierarchical representation with broad potential applications to DAG identification problems.

\bibliographystyle{elsarticle-num}  
\bibliography{refs}  
\end{document}